\documentclass
[showpacs,preprintnumbers,prb,twocolumn,byrevtex,letterpaper,groupedaddress,10pt]{revtex4}%
\usepackage{amsmath}
\usepackage{amssymb}
\usepackage{graphicx}
\usepackage{dcolumn}
\usepackage{bm}
\usepackage{acronym}
\usepackage{amsfonts}%
\begin{document}
\preprint{RYKOV Alexandre / April 18 (2008)}
\title{Vibrational anisotropy and quadrupole interactions of Fe substituted into Mn
site of the charge and orbitally ordered and disordered layered manganites
LnBaMn$_{1.96}$Fe$_{0.04}$O$_{5}$ and LnBaMn$_{1.96}$Fe$_{0.04}$O$_{6}$ (Ln=Y,
Gd, Sm, La, etc.)}
\author{A. I. Rykov,$^{1,2}$ Y. Ueda$^{3}$, K.~Nomura$^{1}$, and M. Seto$^{4}$}
\affiliation{$^{1}$The University of Tokyo, Hongo 7-3-1, 113-8656, Japan, $^{2}$Technology
Crystals Laboratory "Tecrys", Institutskaya 4/1, 630090, Novosibirsk, Russia,
$^{3}$Institute for Solid State Physics, University of Tokyo, 5-1-5,
Kashiwanoha, Chiba 277-8581, Japan, $^{4}$Research Reactor Institute, Kyoto
University, Noda, Kumatori-machi, Osaka 590-0494, Japan}
\date{\today }
\altaffiliation{Corresponding author, rykov3@yahoo.com}

\begin{abstract}
A-site ordered manganites LnBaMn$_{1.96}$Fe$_{0.04}$O$_{5}$
and\ LnBaMn$_{1.96}$Fe$_{0.04}$O$_{6}$ are investigated by x-ray full-profile
diffraction and M\"{o}ssbauer spectroscopy. Powder samples were oriented with
preferred orientation of platy crystallites in the plane of sample surface.
March-Dollase function of preferred orientation was employed in analysing both
the Rietveld patterns and the M\"{o}ssbauer spectra. Combined effects of
preffered orientation and vibrational anisotropy on the line area asymmetry of
M\"{o}ssbauer doublet are analysed. Constructive and destructive interference
between the effects of texture and vibrational anisotropy is observed in
LnBaMn$_{1.96}$Fe$_{0.04}$O$_{6}$ and LnBaMn$_{1.96}$Fe$_{0.04}$O$_{5}$,
respectively. Both series of the manganites show the main axis of electric
field gradient perpendicular to layers (V$_{zz}\parallel c$) \ with $V_{zz}>0$
in oxygen-poor series and $V_{zz}<0$ in oxygen-rich series. Charge-orbital
order (COO) melting around Fe dopants explains the single-site spectra
observed for several Ln in both "O$_{5}$" and "O$_{6}$" series, except
LaBaMn$_{1.96}$Fe$_{0.04}$O$_{5}$. However, the short-range COO persists to be
observed in magnetization and in x-ray patterns.

\end{abstract}
\maketitle

\makeatletter\global\@specialpagefalse

\let\@evenhead\@oddhead

\let\@evenfoot\@oddfoot \makeatother

\section{Introduction}

\label{sec:intro}The manganites LnBaMn$_{2}$O$_{y}$ present a novel class of
layered materilas, in which the layered arrangements of Y and Ba cations
results into the regular architectures of the half-occupied e$_{g}$-orbitals
of Mn$^{3+}$ with the out-of-plane and in-plane orientations for $y=5$ and
$y=6$, respectively. Both series enclose manganese in state of half-doping
mixed valence. The orbital order is coupled to charge order, consisting in the
alternations Mn$^{3+}$/Mn$^{2+}$ and Mn$^{3+}$/Mn$^{4+}$ for $y=5$ and $y=6$,
respectively, along three axes, which are Cartesian for the majority of Ln's,
but slightly oblique in monoclinic\cite{NKIOYU} or triclinic\cite{WAR}
YBaMn$_{2}$O$_{6}$.

Owing to the layered structure the polycrystalline materials are highly
susceptible to preferred orientation withstanding a simple quantitative
description, except the special case of YBaMn$_{2}$O$_{6}$. Recently, such a
quantitative description was suggested\cite{EPL} to be useful in vibrational
spectroscopy on powders of anisotropic materials. As a matter of fact, the
oriented polycrystals consisting of platy crystallites can replace the single
crystals in studies of a variety of anisotropic properties of materials, such
as electric, magnetic or vibrational. Especially, when the polycrystalline
material is a ferromagnet or a superconductor it can be subjected to a
thorough texturing in an external magnetic field.%

\begin{figure}
[ptb]
\begin{center}
\includegraphics[
height=1.8715in,
width=3.2318in
]%
{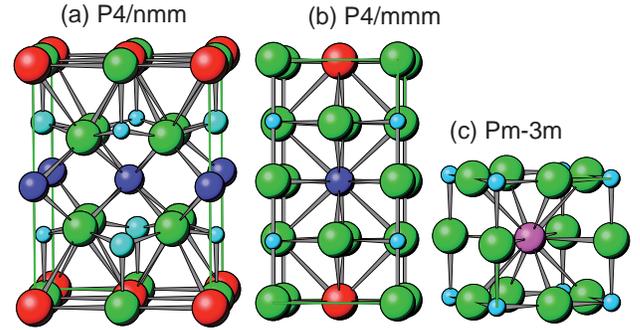}%
\caption{The crystal structures and symmetry groups of the layer-ordered
LnBaMn$_{2}$O$_{5}$(a), LnBaMn$_{2}$O$_{6}$(b) and disordered Ln$_{0.5}%
$Ba$_{0.5}$MnO$_{3}$ (c).}%
\label{f1}%
\end{center}
\end{figure}

In this work, members of both the "O$_{5}$"$~$\cite{MCDRS,MA,Millange} and
"O$_{6}$"$~$ \cite{NKYOU,WAR,NYU,NKIOYU,Aka,NKU} families with Ln=Y, Gd, Sm,
Nd, Pr, La and Ln=Sm$_{1-x}$Nd$_{x}$ solid solutions were doped with 2\% of
$^{57}$Fe and investigated using Rietveld analysis, magnetization and
M\"{o}ssbauer spectroscopy on both random and oriented powders. We investigate
the effect of the Fe substituents on the phase transitions known to occur in
undoped systems. Clean samples exhibit the COO on the long-range scale
observed previously via neutron diffraction\cite{NKYOU,NKIOYU,NYU,Aka,WAR}. We
show that the substitution smears some of the transitions and lowers the
temperature for others.

In charge-ordered state, the Fe substituents display only one component
Fe$^{3+}$ in M\"{o}ssbauer spectra. With doping the stable-valence ion
Fe$^{3+}$ into the mixed-valence site of Jahn-Teller ions, such as Mn$^{2+}%
$/Mn$^{3+}$or Mn$^{3+}/$Mn$^{4+}$, the long-range COO becomes the subject of
suppression by quenched disorder. The short range order is understood to
preserve a favorite arrangements of Mn electronic configurations around the
$^{57}$Fe impurity dopants.

Via the quadrupole interactions and isomer shifts we explore the variety of
structural and valence states adopted by the impurity. In addition, due to the
layered structure the manifestations of vibrational anisotropy appears in
M\"{o}ssbauer line intensities, similarly to layered cuprates, which showed a
notable Goldanskii-Karyagin effect (GKE)\cite{RCR}. In samples with preferred
orientation, the texture effects are combined in M\"{o}ssbauer spectra
intensities with GKE and a technique is suggested to disentangle these
effects. The same technique can be useful to determine the anisotropy in
tensor material properties starting from the data obtained on the aligned
powders. We illustrate the application of this technique to one of such
properties, when the measurements is feasible again with the radiation of the
same wavelength as in M\"{o}ssbauer spectroscopy, however, realizable only on
a large-scale synchrotron facility. This property is the anisotropic
multicomponent phonon density of states (DOS), which request the single
crystals for ordinary measurement\cite{KCR}, however, this work presents a
proposal for determination of both DOS components using the aligned powders
instead of a single crystal.

\section{Experimental details}

The layered A--site-ordered oxygen-saturated manganites LnBaMn$_{1.96}%
$Fe$_{0.04}$O$_{6}$ were prepared for Ln=Y, Gd, Sm, (Sm$_{0.9}$Nd$_{0.1}$),
(Sm$_{0.1}$Nd$_{0.9}$), Nd, Pr and La. The oxides Ln$_{2}$O$_{3}$ and Fe$_{2}%
$O$_{3}$ were mixed with carbonates BaCO$_{3}$ and MnCO$_{3}$. These mixtures
were first annealed in 6N pure Ar (99.9999\%) flow at 1350$^{o}$C. This
annealing has led to obtaining the oxygen-depleted phases LnBaMn$_{1.96}%
$Fe$_{0.04}$O$_{5}$. The synthesis of the oxygen-saturated "O$_{6}$" manganite
is a two-step process as described previously\cite{NKU}. The second step
consisted of annealing in O$_{2}$ flow at low temperature (500$^{o}$C). While
in the first step of the\ oxygen-depletion the layered structure is formed of
the alternating BaO and Y sheets interleaved by the MnO$_{2}$ 'checkerboards',
in the second step this structure is intercalated into Y-layer with the
additional oxygen. A third family of manganites with the same cationic
compositions was obtained in air (without layer-growing treatment in 6N Ar)
and showed a simple perovskite-like disordered structure.%

\begin{figure}
[ptb]
\begin{center}
\includegraphics[
height=3.8207in,
width=3.243in
]%
{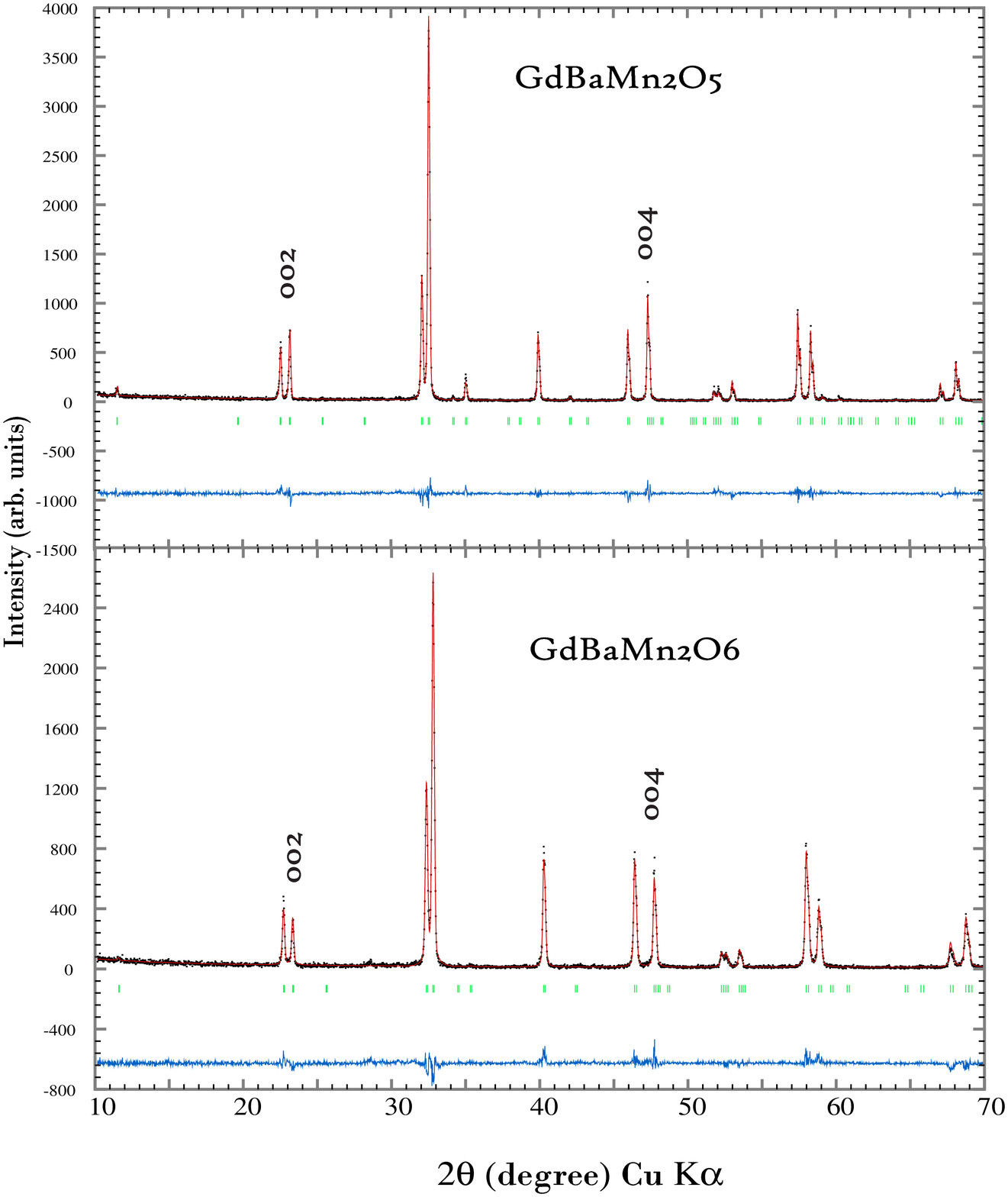}%
\caption{Powder x-ray diffraction patterns of the tetragonal phases of
GdBaMn$_{1.96}$Fe$_{0.04}$O$_{5}$ and GdBaMn$_{1.96}$Fe$_{0.04}$O$_{6}$
refined using space groups $P4/nmm$ (No. 129) and $P4/mmm$ (No.123),
respectively.}%
\label{f2}%
\end{center}
\end{figure}

Rietveld profiles for both "O$_{5}$" and "O$_{6}$" families of layer-ordered
manganites were obtained by means of a "Mac Science" diffractometer using
Cu-$K_{\alpha}$ radiation ($\lambda$= 0.15405 nm\ and 0.15443 nm).
M\"{o}ssbauer spectra were measured at room temperature. Isomer shifts are
referred relatively $\alpha-$Fe. To prepare the oriented powder samples
several methods were used. The powders were wet-spread in alcohol and
dry-spread onto scotch tape or onto blotting-paper. The data were obtained for
powders oriented with their largest dimension in the plane-of the tape with a
marked texture. Rietveld refinements for all diffraction profiles were carried
out using FULLPROF program\cite{DBW,Full}.

Measurements of magnetization were performed using a SQUID magnetometer in an
applied field of 1 kOe at heating the samples from 5 K to $T_{\text{max}} $
and then at cooling from $T=T_{\text{max}}$ down to 5 K. This measurment
protocol was applied in LnBaMn$_{1.96}$Fe$_{0.04}$O$_{6}$ for Ln=Sm with
$T_{\text{max}}$=400 K and for Ln=(Nd$_{0.9}$Sm$_{0.1}$) with $T_{\text{max}%
}=$370 K. The magnetization in YBaMn$_{1.96}$Fe$_{0.04}$O$_{6}$ was measured
first at heating from ambient temperature to $T_{\text{max}}$= 600 K (Ln=Y)
and then at cooling from $T=T_{\text{max}}$ down to 5 K. The sample was then
remagnetized at 5K by setting the external field $H=0\pm0.01$ kOe followed by
reapplying $H=1$ kOe. Finally, this sample magnetization was measured at
heating up to 300 K.

\section{Results and discussion}

\subsection{Structural considerations.}

\subsubsection{Room-temperature lattice parameters and symmetry groups}

Prior to first reports on the manganites YBaMn$_{2}$O$_{5}$%
\cite{MCDRS,MA,MSCB} and LaBaMn$_{2}$O$_{y}$ ($y=5$ and $6$%
)\cite{Millange,Caignaert} which crystallized in the perovskite-like bilayered
tetragonal structures, the ferrocuprate YBaCuFeO$_{5}$ with the same structure
built up of bipyramidal layers was already known\cite{Er} since 1988. On the
other hand, the oxygen-rich ($y=6$) layered manganite system had no prototype
among cuprates. There exist two important features of the manganite systems,
distinguishing them from the cuprate one. First, the layers of Ba and Ln
develop in LaBaMn$_{2}$O$_{y}$ only if the system is deoxygenated ($y\sim5$)
at high-temperature synthesis. Excess of oxygen during the high-temperature
annealing breaks the layered structure and create the disordered isotropic
phase. Second, both the "O$_{5}$" and "O$_{6}$" structures exhibit the
in-plane ordering mixed valence ions Mn$^{2+}$/Mn$^{_{3+}}$ and Mn$^{3+}%
$/Mn$^{4+}$, respectively, both associated with COO. We observed that all of
our x-ray profiles from "O$_{5}$" samples are perfectly fitted with the
$P4/nmm$ model (Fig.1). The quality of fit was strongly declined when we
attempted to fit the patterns in terms of $P4/mmm$ model. This result suggests
that the COO is essentially preserved in our samples having the Fe ions placed
into 2\% of the Mn sites. However, from M\"{o}ssbauer single-site spectra a
COO meltdown is suggested, as shown below. Such a meltdown is not denotiative
of full randomness. Short range order in the arrangement of charges and
orbitals is preserved. Indeed, since no superstructure reflections were
observed, associated with lowering symmetry from $P4/mmm$ to $P4/nmm$, our
Rietveld refinement preference for the symmetry group $P4/nmm$ indicates only
that the COO is kept on a short range. In pure LaBaMn$_{2}$O$_{5},$ the
supercell reflections are observable with electron diffraction\cite{Millange},
while Fe-doping may broaden them and suppress their intensity to the level
insufficient for observation. The coexistence of the preserved short-range COO
with the doping-induced quenched disorder is discussed below along with
M\"{o}ssbauer spectra.%

\begin{figure}
[ptb]
\begin{center}
\includegraphics[
natheight=3.646900in,
natwidth=4.581800in,
height=2.8504in,
width=3.576in
]%
{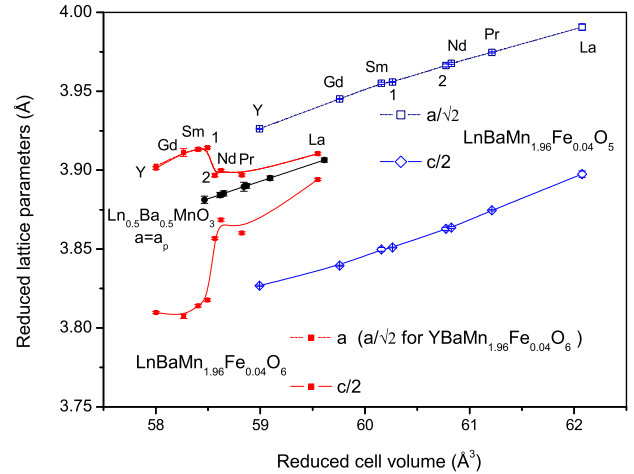}%
\caption{Lattice parameters of the reduced perovskite-like cell vs. volume of
this cell in 2\% Fe-doped manganites. Mixed-rare-earths manganites Sm$_{0.9}%
$Nd$_{0.1}$BaMn$_{1.96}$Fe$_{0.04}$O$_{y}$ and Sm$_{0.1}$Nd$_{0.9}%
$BaMn$_{1.96}$Fe$_{0.04}$O$_{y}$ are denoted by "1" and "2", respectively.}%
\label{f3}%
\end{center}
\end{figure}

In contrast to the singular "O$_{5}$"-phase ($P4/nmm-$phase$)$, that is common
for all the rare-earth elements, a few different modifications are known for
the oxygen-saturated "O$_{6}$" A-site ordered layered manganites, depending on
the size of Ln \cite{NKU,NKYOU,NKIOYU,WAR}. All of them are built up of
octahedral bilayers composed of octahedra distorted in one way or other. Our
Rietveld refinements were carried out using several symmetry groups and
structure models known from the literature\cite{NKU,NKYOU,NKIOYU,WAR}. These
treatments have shown that the tetragonal cell $a_{\text{p \ }}\times$
\ $a_{\text{p \ }}$\ \ $\times$\ \ \ $2a_{\text{p \ }}$suits best (Fig.2) to
all the samples in the layer-ordered oxygen-saturated family, with one
exception of YBaMn$_{1.96}$Fe$_{0.04}$O$_{6}$. The structure of the latter was
refined using the monoclinic symmetry (space group P2, No.3).%

\begin{figure}
[ptb]
\begin{center}
\includegraphics[
natheight=5.110200in,
natwidth=3.659000in,
height=4.83in,
width=3.4705in
]%
{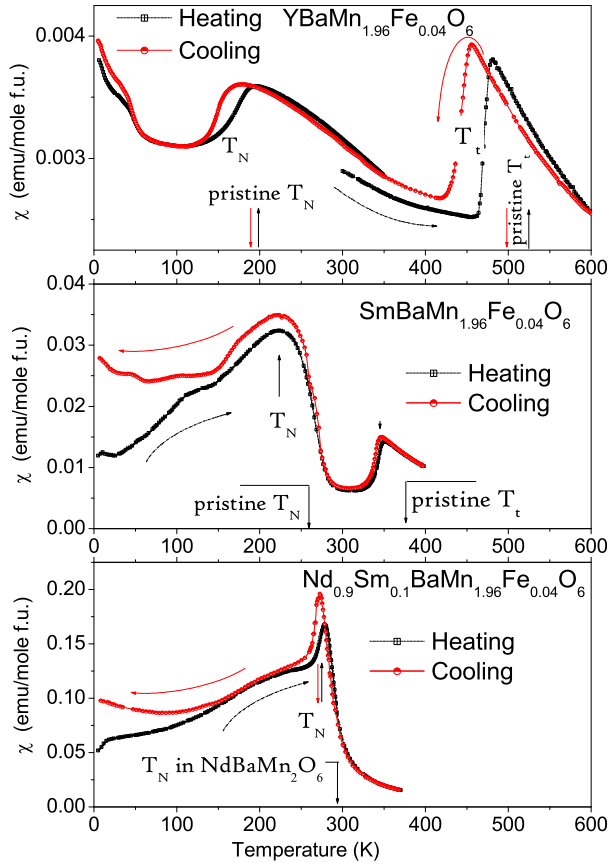}%
\caption{Magnetic susceptibility $M/H$ measured in the external field $H$ of 1
kOe per mole of formula units in LnBaMn$_{1.96}$Fe$_{0.04}$O$_{6}$ for Ln=Y,
Sm and (Nd$_{0.9}$Sm$_{0.1}$). The zero-fied-cooled magnetization was measured
at heating the samples up to $T_{\text{max}}$ of 600 K (Ln=Y), 400 K (Ln=Sm)
and 370 K (Ln=Nd$_{0.9}$Sm$_{0.1}$) and then at cooling from $T=T_{\text{max}%
}$. The arrows indicate the temperatures of phase transitions in corresponding
pure LnBaMn$_{2}$O$_{6}$ without Fe substitution as reported
previously\cite{NYU,NKIOYU}.}%
\label{f4}%
\end{center}
\end{figure}

To smmarize, the lattice cells were fitted with the volume double or quadruple
of perovskite one ($V\cong2a_{\text{p}}^{3}$ or 4$a_{\text{p}}^{3} $) owing to
the lattice parameters relations:%

\begin{tabular}
[c]{l}%
$a=b\cong\sqrt{2}a_{\text{p \ }}\text{; \ }c\cong2a_{\text{p \ \ \ }}\text{for
LnBaMn}_{1.96}\text{Fe}_{0.04}\text{O}_{5}$\\
$a=b\cong a_{\text{p \ }}\text{; \ \ \ \ \ }c\cong2a_{\text{p \ \ \ }%
}\text{for LnBaMn}_{1.96}\text{Fe}_{0.04}\text{O}_{6}$\\
$a=a_{\text{p \ }}$ \ \ \ \ \ \ \ \ \ \ \ \ \ \ \ \ \ for $\text{Ln}%
_{0.5}\text{Ba}_{0.5}\text{Mn}_{0.98}\text{Fe}_{0.04}\text{O}_{3}$%
\end{tabular}

\subsubsection{Distortions of perovskite cell}

Comparing the parameters of reduced lattice cell (i.e. a distorted perovskite
cell) one can analyze the extent of distortion depending on Ln, in function of
the volume of the reduced cell. Fig. 3 summarizes the results of refinement of
the lattice cell parameters. The reduced parameters $a/\surd2$ ($a/2$) and
$c/2$ are plotted against the reduced cell volume $V/4 $ ($V/2$). From Y to La
the volume $V$ varies in LnBaMn$_{1.96}$Fe$_{0.04}$O$_{5}$ within 5\% and both
$a$ and $c$ vary in the range of 1.7\%. Distortion of the reduced cell can be
calculated in LnBaMn$_{2}$O$_{5}$ as $D_{5}=2(a/\surd2-c/2)/(a/\surd2+c/2)$ .
From Y to La the distortion decreases from 2.56\% to 2.36\%, respectively. In
the oxygen-saturated series, the distortion $D_{6}=2(a-c/2)/(a+c/2)$ does not
show a stationary level from Y to La, but changes abruptly between Sm and Nd.

\subsubsection{Phase transitions}

We observe in Fig. 4 that there occurs in LnBaMn$_{1.96}$Fe$_{0.04}$O$_{6}$
the phase transitions associated with COO and magnetic ordering. The same
jumps and humps of susceptibility are observed as in the undoped LnBaMn$_{2}%
$O$_{6}$, however, the corresponding transition temperatures are suppressed.
According to size of Ln three groops of the oxygen-saturated manganites were
specified previously\cite{JPSJ2002} as Ln(I) $=$(Y, Tb, Dy, Ho), Ln(II)=(Sm,
Eu, Gd) and Ln(III)=(Nd, Pr, La). A member from each of these families was
investigated in this work for the effect of the Fe substituents on the
transitions manifested in magnetic properties.

The transition sequences (temperatures $T_{\text{t}}$ , $T_{\text{N}}$ ) in
the lightly (i.e. 2\%) Fe-doped manganites are essentially same as in undoped
manganites, but differ in details. In YBaMn$_{1.96}$Fe$_{0.04}$O$_{6}$ and
SmBaMn$_{1.96}$Fe$_{0.04}$O$_{6}$, the highest in temperature jump of
magnetzation is associated with a structural transition, which is
monoclinic-to-triclinic\cite{NKIOYU} or triclinic-to-monclinic\cite{WAR} for
Ln=Y, and tetragonal-to-orthorhombic\cite{Akah} for Ln=Sm. Orbital ordering is
now believed\cite{UN} to accompany these structural transition ($T=T_{\text{t}%
}$), while complete charge ordering is attributed\cite{NKU,NKIOYU}, e.g. for
Ln=Y, to a separate small hump shifted from $T_{\text{t}}$ to lower
temperature by $\Delta_{\text{t}}\approx40$ K. In the undoped manganites, the
temperature $T_{\text{CO}}=T_{\text{t}}-\Delta_{\text{t}}$ is associated with
sharp localization of charge carriers. In the magnetization of YBaMn$_{1.96}%
$Fe$_{0.04}$O$_{6}$, the large jump is observed apparently without any small
foregoing hump. Interestingly, similar disappearance of the hump in
magnetization caused by Fe doping was reported for orbital ordering transition
in BiMnO$_{3}$\cite{Belik}. Smearing the transition over a broad temperature
range induced by Fe substitution suppresses such a hump.

Another key feature of the doped systems stems from the fact that the values
of $T_{\text{t}}$ are slightly suppressed compared to undoped YBaMn$_{2}%
$O$_{6}$ and SmBaMn$_{2}$O$_{6}$\cite{NKIOYU,NYU}. The suppression ranges are
$\Delta T_{\text{t}}($2\%Fe$)=$50K and 40 K for Ln=Y and Sm, respectively.

In contrast to similar values of $\Delta T_{\text{t}}$, there occurs a large
difference between the cases of Ln=Y and Sm for the shift of the transition
temperature with the reversal of temperature sweep direction. Such a shift is
associated with an energy barrier for nucleation of a new phase within the
region of overheating or undercooling the preceding phase. The hysteresis
indicates strongly the first-order character of transition that was observed
also in undoped YBaMn$_{2}$O$_{6}$\cite{NKIOYU}. The large nucleation barrier
is observed in YBaMn$_{2}$O$_{6}$ but not in SmBaMn$_{2}$O$_{6}$. This is in
agreement with a very small structural distortion in SmBaMn$_{2}$O$_{6}$ at
$T_{\text{t}}$ as reported by Akahoshi et al\cite{Akah}.

Temperatures of Neel ($T_{\text{N}}$) are also notably suppressed in both
cases, as well as in the third group member, Nd$_{0.9}$Sm$_{0.1}$BaMn$_{1.96}
$Fe$_{0.04}$O$_{6}$. The antiferromagnetic transitions humps were
observed\cite{NKYOU,NYU} at 290 K and 250 K in undoped NdBaMn$_{2}$O$_{6}$ and
SmBaMn$_{2}$O$_{6}$, respectively, therefore, the $T_{\text{N}}$\ value of 286
K is expected for the solid solution Nd$_{0.9}$Sm$_{0.1}$BaMn$_{2}$O$_{6}$.
Remaining suppression $\Delta T_{\text{N}}$ $\simeq10$ K should be attributed
to the effect of Fe substitution. Temperature ranges for sweep-reversal
hysteresis around $T_{\text{N}}$ are not much different between the three.

\subsubsection{Preferred orientation}

In both series of the layered manganites the samples displayed some degree of
preferred orientation which is directly seen in x-ray diffraction patterns
through an enhancement of the reflections of the type $00l$ (Fig.2). This
indicates the platy habits of thin crystallites with the easy cleavage planes
parallel to the layers. Table 1 summarizes the results of refinement of the
preferred orientation parameters.

\textbf{Table 1. } Parameters of March-Dollase function, $G_{1}$ and $G_{2}$,
Eq.(1), refined from x-ray diffraction patterns.%

\begin{tabular}
[c]{l|llllll|}\cline{2-7}
& \multicolumn{3}{|l}{LnBaMn$_{1.96}$Fe$_{0.04}$O$_{5}$} &
\multicolumn{3}{|l|}{LnBaMn$_{1.96}$Fe$_{0.04}$O$_{6}$}\\\hline
\multicolumn{1}{|l|}{Ln} & G$_{1}$ & \multicolumn{1}{|l}{G$_{2}$} &
\multicolumn{1}{|l}{$\frac{M(0)}{M(\pi/2)}$} & \multicolumn{1}{|l}{G$_{1}$} &
\multicolumn{1}{|l}{G$_{2}$} & \multicolumn{1}{|l|}{$\frac{M(0)}{M(\pi/2)} $%
}\\\hline
\multicolumn{1}{|l|}{Y} & 0.61(3) & 0.27(2) & 5.64 & N/A*** & N/A*** &
---\\\cline{1-1}%
\multicolumn{1}{|l|}{Gd} & 0.61(1) & 0.47(2) & 3.88 & 0.57(3) & 0.84(2) &
1.88\\\cline{1-1}%
\multicolumn{1}{|l|}{Sm} & 0.73(1) & 0.68(4) & 1.71 & 0.76(3) & 0.78(2) &
1.38\\\cline{1-1}%
\multicolumn{1}{|l|}{1*} & 0.61(2) & 0.74(3) & 2.18 & 0.74(3) & 0.76(2) &
1.48\\\cline{1-1}%
\multicolumn{1}{|l|}{2**} & 0.80(2) & 0.59(10) & 1.57 & 0.97(5) & 0.99(1) &
1.00\\\cline{1-1}%
\multicolumn{1}{|l|}{Nd} & 0.47(4) & 0.75(5) & 3.80 & 0.55(10) & 0.94(3) &
1.35\\\cline{1-1}%
\multicolumn{1}{|l|}{Pr} & 0.45(6) & 0.57(10) & 7.56 & 0.46(22) & 0.97(2) &
1.31\\\cline{1-1}%
\multicolumn{1}{|l|}{La} & 0.68(2) & 0.75(15) & 1.74 & 0.96(3) & 0.86(3) &
1.03\\\hline
\end{tabular}

*1: Sm$_{0.9}$Nd$_{0.1}$BaMn$_{1.96}$Fe$_{0.04}$O$_{x};$

**2: Sm$_{0.1}$Nd$_{0.9}$BaMn$_{1.96}$Fe$_{0.04}$O$_{x}$

***Refinement with the uniaxial texture along [$00l$] was not applicable by
reason of monoclinic structure.\bigskip

The March-Dollase (MD)\ function\cite{inc} depends on three variables: the
texture-axis-misfit angle $\theta$ and \ two profile-refinable parameters
$G_{1}$, $G_{2}$ (Ref.\onlinecite {Dollase}),%

\begin{equation}
M(\theta,G_{1},G_{2})=G_{2}+\frac{1-G_{2}}{\left[  (G_{1}\cos\theta)^{2}%
+G_{1}^{-1}\sin\theta\right]  ^{\frac{3}{2}}} \label{1}%
\end{equation}
That is to say, $\theta$ is the acute angle between the x-ray scattering
vector $h\mathbf{a}^{\ast}+k\mathbf{b}^{\ast}+l\mathbf{c}^{\ast}$ and the axis
of preferred orientation. In powders of single-crystalline particles, the
preferred axis is dictated by the grain shape. The cylindrical-symmetry
texture axis runs along the whisker-like or fiber-like crystals, but along the
\textit{normal} to the plate-like crystals. We fitted all our x-ray patterns
with the uniaxial texture along the direction [$00l$]. The Eq.(\ref{1})
describes the density of poles, which come into reflection position at a given
Bragg angle $\theta=\theta_{hkl}$. From the viewpoint of conserving the
scattering matter the function $M(\theta)$ is a true distribution function
suitable for quantitative analysis. In this sense, the March-Dollase
distribution is much better than Gaussian distribution, originally suggested
by Rietveld\cite{Riet}. The parameter $G_{2}$ desrcibes the fraction of the
sample that is not textured and the parameter $G_{1}$ describes the degree of
alignment within the textured fraction. The diffraction intensities are scaled
by the distribution $M(\theta)$ which culminates at $\theta=0$ for $G_{1}<1$ ,
but at $\theta=\pi/2$ for $G_{1}>1$. Therefore, both plate-like and
needle-like powders are refinable with the universal function $M(\theta)$. The
domains $0<G_{1}<1$ and $1<G_{1}<\infty$ correspond to the platy and acicular
habits, respectively, while fully isotropic powder fits to $G_{1}=1$.\ 

Two parameters of the textured sample anisotropy $G_{1}$ and $G_{2}$ presents
the full set of texturing characteristics, however, it is convenient to
simplify the sample comparison using the single parameter $M(0)/M(\pi/2)$
given by a combination of $G_{1}$ and $G_{2}$. This ratio displays directly
the enhancement of right-hand line within the closely-spaced pairs
$I(002)/I(100)$ and $I(004)/I(002)$ (Fig.2). From the Table 1 it becomes clear
that LnBaMn$_{1.96}$Fe$_{0.04}$O$_{5}$ shows much stronger tendency to
texturing than LnBaMn$_{1.96}$Fe$_{0.04}$O$_{6}$. This result could be
explained by variations in crystallite size and aspect ratio.%
\begin{figure}
[ptb]
\begin{center}
\includegraphics[
natheight=7.810100in,
natwidth=5.802000in,
height=4.0767in,
width=3.0381in
]%
{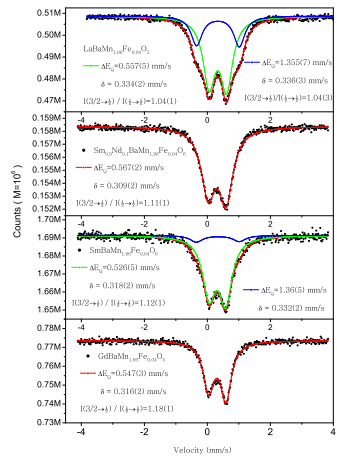}%
\caption{Mossbauer spectra in oriented samples of $^{57}$Fe-doped oxygen-poor
series of the A-site ordered layered manganites LnBaMn$_{1.96}$Fe$_{0.04}%
$O$_{5}$. Fitted values of isomer shift, quadrupole splitting, and area
asymmetry are indicated on the spectra. The spectra are fitted with one or two
asymmetric doublets. Meltdown of Mn$^{2+}$/Mn$^{3+}$charge order results in
single-site spectra for $^{57}$Fe, while residue of the second doublet is
attributed to the remainder of the charge/orbital order. }%
\label{f5}%
\end{center}
\end{figure}

Returning to the preparation conditions of these series of samples, one sees
that LnBaMn$_{1.96}$Fe$_{0.04}$O$_{6}$ was prepared from LnBaMn$_{1.96}%
$Fe$_{0.04}$O$_{5}$ in oxygen at the temperature as low as 500$^{o}$C. In such
a low temperature process, the crystallites could be destroyed and diminished
in size, but hardly could grow. It follows indeed from Fig.3 that the large
compression of the lattice takes place at oxygenation. It was observed
previously in GdBaMn$_{2}$O$_{5+x}$ \cite{Taskin}, that the rate of oxygen
intercalation into the layered oxide is extremely high. Strains associated
with the oxygenation are very large. The crystal cracking and breakup in
lateral dimension are likely to accompany the oxygen intercalation. The
crystal cracks and strains observed during oxygenation were attributed to
orthorhombicity\cite{Taskin}. We have stabilized the orthorhombic phase
GdBaMn$_{2}$O$_{5.5}$ using a more complicated sequence of thermal
treatments\cite{parallel}. The orthorhombic distortion in GdBaMn$_{2}$%
O$_{5.5}$ is indeed much larger than either D$_{5}$ or D$_{6}$, calculated
above. Large distortion arises from channel-like structure of the half-filled
layer LnO$_{0.5}$ and this could be the plausible origin of the crystal
dispergating with oxygenation\cite{parallel}.

\subsection{M\"{o}ssbauer spectroscopy}

\subsubsection{Oxygen-depleted phase LnBaMn$_{1.96}$Fe$_{0.04}$O$_{5}$}

M\"{o}ssbauer spectra in oriented samples of LnBaMn$_{1.96}$Fe$_{0.04}$O$_{5}$
are clearly asymmetric. In the samples with Ln=Gd and Ln=(Sm$_{0.9}$Nd$_{0.1}%
$) the spectra are fitted with single asymmetric doublet. However, two
doublets are crucial to fit the spectra for Ln=La and Sm (Fig.5).

Spectral asymmetry increases from La towards Gd and this increase correlates
with increasing distortion D$_{5}$. The second doublet is not enough resolved
to fit separately its asymmetry. Therefore, its asymmetry was fixed at fitting
to be the same as the asymmetry of the major doublet.

As the size of Ln decreases we observe the systematic decrease in the area of
second doublet. An exception from this series is made by the mixed rare-earth
manganite Sm$_{0.9}$Nd$_{0.1}$BaMn$_{1.96}$Fe$_{0.04}$O$_{5}$ showing a single
doublet, although the average size of (Sm$_{0.9}$Nd$_{0.1}$) is larger than
the size of Sm, while SmBaMn$_{1.96}$Fe$_{0.04}$O$_{5}$ still shows a presence
of the second doublet. However, this observation is in line with our
interpretation of the single-site spectrum as originating from the meltdown of
the charge order induced by the pointlike quenched disorder. Indeed, the
additional pointlike disorder in Sm$_{0.9}$Nd$_{0.1}$BaMn$_{1.96}$Fe$_{0.04}%
$O$_{5}$ is related to disorder in the Ln site.%

\begin{figure}
[ptb]
\begin{center}
\includegraphics[
height=2.8291in,
width=3.538in
]%
{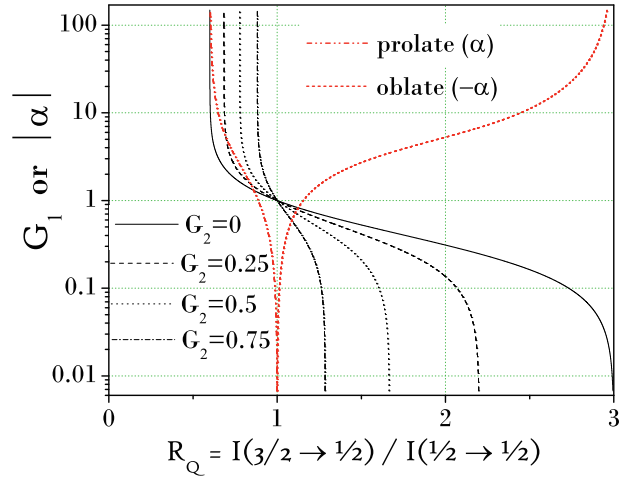}%
\caption{Determination of the parameters of texture $G_{1}$ and vibrational
anisotropy $\alpha$ starting from the ratio of line intensities of
M\"{o}ssbauer doublet. Either texture or GKE is supposed to be a single source
of spectral asymmetry. Strength of preferred orientation $G_{1}$ is plotted
vs. spectral asymmetry $R_{Q}=I_{\pm1/2\rightarrow\pm3/2}/I_{\pm
1/2\rightarrow\pm1/2}$ for four values of the fraction of unoriented phase
$G_{2}$.}%
\label{f6}%
\end{center}
\end{figure}

\subsubsection{Coexistense of the short-range COO with quenched disorder}

An important result appears in the fact that Fe sees the charge order broken
more easily for smaller Ln by the same level of doping. The double-site
spectrum in LaBaMn$_{1.96}$Fe$_{0.04}$O$_{5}$ evidences that the charge order
is robust for Ln=La, so that the Fe species ocuppy both Mn(II) and Mn (III)
sites, although the area ratio is 7:3 instead of 5:5 expected for random
occupation. Such a random occupation might be expected only in case of
long-range charge order. However, iron can adopt more easily its preferred
site when the charge and orbital ordered domains decrease in size.

There occurs enough charge fluidity in GdBaMn$_{1.96}$Fe$_{0.04}$O$_{5}$ and
in Sm$_{0.9}$Nd$_{0.1}$BaMn$_{1.96}$Fe$_{0.04}$O$_{5}$ to realize the unique
surrounding\ for overwhelming majority of dopants. On the other hand, in
LaBaMn$_{1.96}$Fe$_{0.04}$O$_{5}$ and in SmBaMn$_{1.96}$Fe$_{0.04}$O$_{5}$, we
observe the robust COO that has a longer correlation lengths, such that a
minor but significant part of dopants remains in a different surrounding.

The second doublet shows the same isomer shift but twice increased quadrupolar
splitting. Both doublets originate from $^{57}$Fe in pyramidal coordination,
however, one of these doublets originates from the site having a larger apical
distance and stronger Jahn-Teller character. Trying to understand the origin
of difference between doping behaviors for large and small Ln's, let us
compare the pyramidal interatomic distances in LaBaMn$_{2}$O$_{5}$ and in
YBaMn$_{2}$O$_{5}$ as reported previously\cite{Millange,MSCB}. The difference
between Mn(II) and Mn(III) pyramids is similar for Ln=La and Y, however, in
YBaMn$_{2}$O$_{5}$, both Mn(II) and Mn(III) pyramids are more elongated in
apical dimension, and compressed in equatorial dimension. Such a distortion
appears thus to facilitate the melting of the charge and orbital order around
the doped Fe species.%

\begin{figure}
[ptb]
\begin{center}
\includegraphics[
natheight=3.715000in,
natwidth=4.833500in,
height=2.3702in,
width=3.0764in
]%
{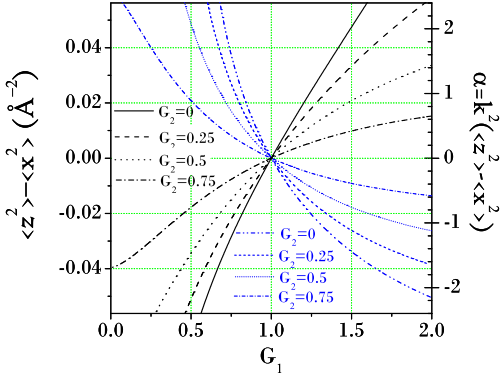}%
\caption{The effective vibrational anisotropy parameters $\alpha$ and
$\alpha/k^{2}$ versus the equivalent degree of alignment $G_{1}\ $at four
values of the unoriented phase $G_{2}$. The curves with positive slope refer
to the true assignment, and the curves with negative slopes refer to the false
assignment of the doublet lines to the $\pm1/2\rightarrow\pm3/2$ and
$\pm1/2\rightarrow\pm1/2$ transitions. }%
\label{f7}%
\end{center}
\end{figure}

\subsubsection{Two sources of the doublet asymmetry}

Known origins of the line area asymmetry in M\"{o}ssbauer spectra are the
preferred orientation of crystallites and vibrational anisotropy. In a random
polycrystalline material, the ordinary single-site spectra in paramagnetic
domain are the symmetric doublets. The equal intensities of both lines result
from random-powder averaging, unless GKE takes place. Preferred orientation
would induce the asymmetry with or without GKE. We discuss first the separate
ocurrence of texture effects and GKE and start from the texture effects.

\paragraph{Texture effects}

Let us consider the doublet intensity ratio for a single crystallite. This
ratio depends on the orientation of the wave vector of the incident x-ray
quantum with respect to the axes of the electric field gradient (EFG) tensor
of the site wherein the $^{57}$Fe nucleus is located. The angular dependence
of Clebsch-Gordan coefficients\cite{Sh} results in the $\theta$-dependent
intensity ratio of the quadrupole doublet $R_{Q}$ for each crystallite:%

\begin{equation}
R_{Q}=\frac{I_{\pm1/2\rightarrow\pm3/2}}{I_{\pm1/2\rightarrow\pm1/2}}%
=\frac{1+\cos^{2}\theta}{2/3+\sin^{2}\theta}\text{ \ \ \ } \label{033}%
\end{equation}

When the unoriented powder is used for measuring the spectra the angular
averaging gives $\left\langle \sin^{2}\theta\right\rangle =2/3$, $\left\langle
\cos^{2}\theta\right\rangle =1/3$ and $R_{Q}=1$. In case of oriented powder,
using the MD texture function (Eq.\ref{1}), we substitute $\left\langle
\sin^{2}\theta\right\rangle $ and $\left\langle \cos^{2}\theta\right\rangle $
$=1-\left\langle \sin^{2}\theta\right\rangle $ with
\begin{equation}
\left\langle \sin^{2}\theta\right\rangle \equiv V(G_{1},G_{2})=\int_{0}%
^{1}M(\theta,G_{1},G_{2})\sin^{3}\theta d\theta\label{07}%
\end{equation}
where%
\begin{align}
V(G_{1},G_{2}) &  =\frac{2}{3}G_{2}+(1-G_{2})v(G_{1})\label{06}\\
v(G_{1}) &  =\frac{G_{1}^{2}}{\varepsilon^{2}(G_{1})}-\frac{\beta(G_{1}%
)}{2\varepsilon^{3}(G_{1})}\\
\varepsilon(G_{1}) &  =\sqrt{G_{1}^{2}-G_{1}^{-1}}\text{ \ \ \ }\\
\beta(G_{1}) &  =\ln(2G_{1}^{3}+2\sqrt{G_{1}^{6}-G_{1}^{3}}-1)
\end{align}
Both factors $\varepsilon(G_{1})$ and $\beta(G_{1})$ are imaginary for
$0<G_{1}<1$, however, $V(G_{1})$ is real in full range $0<G_{1}<\infty$. 

Rietveld analysis allows one to fit the parameters of the degree of alignment
$G_{1}$ and the aligned fraction $G_{2}$ starting from a set of x-ray
diffraction intensities. A similar problem can be formulated in Mossbauer
spectroscopy: to determine $G_{1}$ and $G_{2}$ starting from the spectral
asymmetry. In the area of small asymmetries, the effect of $G_{1}$ and $G_{2}$
on the asymmetry turns out to be nearly equal, as will be shown below. In
Fig.6, the degree of alignment $G_{1}$ is plotted vs. the asymmetry $R_{Q}$
for several values of the textured fraction $G_{2}$.\bigskip

\paragraph{Vibrational anisotropy}

The vibrational anisotropy $\alpha$ is also plotted in Fig.6 vs. $R_{Q}$. It
is expressed through the absolute value of the wave vector for $\gamma
$-radiation and the mean-square vibrational displacements along $V_{zz}$, and
in perpendicular direction, $\alpha=k^{2}(\langle z^{2}\rangle-\langle
x^{2}\rangle)$. This plot was calculated following the integral of each line
intensity scaled with the angle-dependent Lamb-M\"{o}ssbauer factor for
uniaxial symmetry $\exp(-\alpha\cos^{2}\theta)$:%

\begin{equation}
R_{Q}(\alpha)=\frac{\int_{0}^{\pi/2}(1+\cos^{2}\theta)e^{-\alpha\cos^{2}%
\theta}d\theta}{\int_{0}^{\pi/2}(2/3+\sin^{2}\theta)e^{-\alpha\cos^{2}\theta
}d\theta}\text{ \ \ \ } \label{22}%
\end{equation}
The Eq.(\ref{22}) is appropriate in randomized powders of all our "O$_{6}$"
phases, except YBaMn$_{1.96}$Fe$_{0.04}$O$_{6}$. The sigmoidal function
$R_{Q}(\alpha)$ is decreasing from 3 to 0.6 when $\alpha$ is varied from
$-\infty$ to $+\infty$. The inverse function $\alpha(R_{Q})$ is shown in Fig.6
as a semilog plot, having two branches. The right-hand branch ($R_{Q}>1 $)
corresponds to oblate, pancake-shaped ellipsoids of atomic thermal
displacements (ATD), and the less-dispersive left-hand branch correspond to
prolate, or cigar-like ATD.

From Fig.6 it becomes clear that the underbalanced line area ratio ($R_{Q}<1)$
is related to either prolate ATD or to acicular crystallite textures. The
overbalanced ratio ($R_{Q}>1$) is produced either by oblate ATD or by platy
crystallite textures. Mixing these effects will be analysed below in \S 8.%

\begin{figure}
[ptb]
\begin{center}
\includegraphics[
natheight=4.847300in,
natwidth=3.632200in,
height=4.4278in,
width=3.3278in
]%
{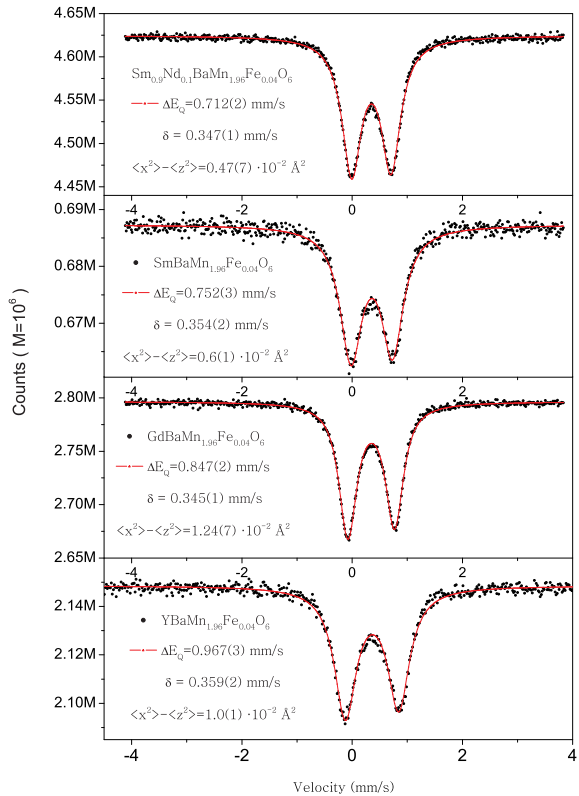}%
\caption{Mossbauer spectra in oriented samples of $^{57}$Fe-doped oxygen-rich
series of the charge and orbitally ordered layered manganites LnBaMn$_{1.96}%
$Fe$_{0.04}$O$_{6}$. The spectra are fitted with one asymmetric doublet.
Meltdown of Mn$^{3+}$/Mn$^{4+}$charge order results in single-site spectra for
$^{57}$Fe. Fitted values of isomer shift, quadrupole splitting, and area
asymmetry are indicated on the spectra. The values of $\langle x^{2}%
\rangle-\langle z^{2}\rangle$ are the "effective" anisotropies given by the
sum of true and equivalent to the pair ($G_{1},G_{2}$) values.}%
\label{f8}%
\end{center}
\end{figure}
\begin{figure}
[ptbptb]
\begin{center}
\includegraphics[
height=4.4884in,
width=3.384in
]%
{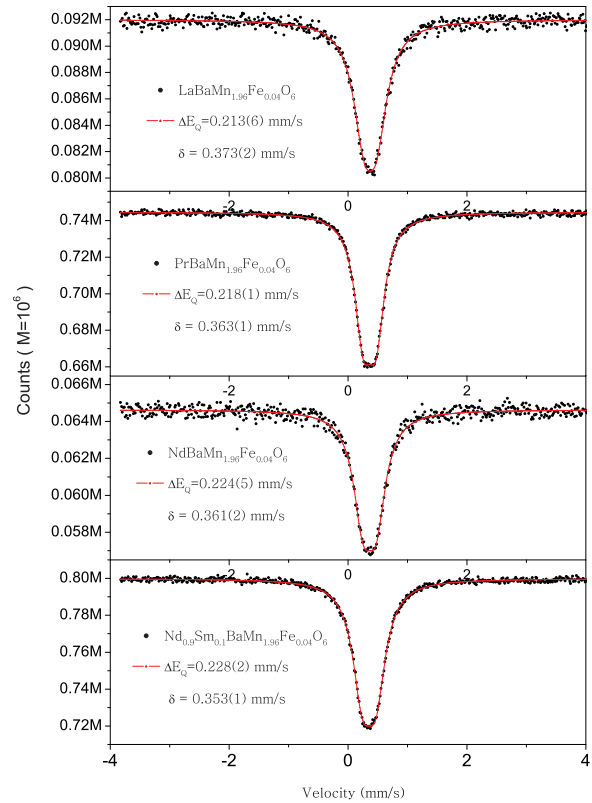}%
\caption{Mossbauer spectra in the large-size-Ln subseries of the oxygen-rich
series of the A-site ordered layered manganites LnBaMn$_{1.96}$Fe$_{0.04}%
$O$_{6}$ showing no COO transition. The spectra are fitted with one symmetric
doublet.}%
\label{f9}%
\end{center}
\end{figure}

\subsubsection{Line assignement}

The case $R_{Q}>1$ refers in Eqs.(\ref{033},\ref{22}) to stronger line
$\pm1/2\rightarrow\pm3/2$ and weaker line $\pm1/2\rightarrow\pm1/2$. The line
$\pm1/2\rightarrow\pm3/2$ lies higher in energy than the line $\pm
1/2\rightarrow\pm1/2$ if $V_{zz}>0$. \textit{Vice versa, }if $V_{zz}<0$, the
lines swap the energy positions.

We attribute the more intense line in Fig. 5 to the transition $\pm
1/2\rightarrow\pm3/2$ assuming that the electric field gradient $V_{zz}>0$ is
oriented along $z-$axis perpendicular to layers. This is in agreement with the
point charge model for pyramidal coordination \cite{HI}. Thus, either oblate
ATD or platy crystallite textures underlie the observed in Fig.5 spectral
asymmetry. Among the reports of neutron diffraction profile refinements in
LnBaMn$_{2}$O$_{5}$, no data are known for an anisotropic ATD because all
profiles were refined using B$_{iso}$\cite{MCDRS,MA,Millange}. The anisotropic
ATD, however, were refined previously for pyramidal coordination of
YBaFeCuO$_{5}$. The oblate ATD ellipsoids were never observed in pyramidal
coordination. On the opposite, the ATD tensor shaped as very extended prolate
'cigar' was found in YBaFeCuO$_{5}$\cite{Mom}. Thus, the self-consistent
combination is $V_{zz}>0$ and prevailing asymmetry owing to platy-crystallite
texture. The asymmetry would be reduced by prolate ATD, but not fully compensated.

The compensation effects can be described quantitatively if we find what
texture parameters are equivalent to vibrational anisotropy parameter in its
effect on asymmetry. Having arrived to the quantitative description of texture
using the Eqs.(\ref{06}-\ref{07}) and Fig.6, we can now plot the vibrational
parameter $\alpha$ vs. the equivalent degree of alignment $G_{1}$(Fig.7).
\ Due to the second parameter G$_{2}$ a family of curves is produced, so that
the effective $\alpha$ corresponding to a pair ($G_{1},G_{2}$) can be found
from the plot.

Knowledge of the EFG orientation and sign is not ubiquitous in M\"{o}ssbauer
spectroscopy. Since the correct attribution of the doublet lines to the
$\pm1/2\rightarrow\pm1/2$ and $\pm1/2\rightarrow\pm3/2$ transitions is not
always obvious, we plot in Fig.7 also the family of curves, which correspond
to the false attribution. True attribution regions correspond to the
bottom-left (oblate-platy) and top-right (prolate-acicular) quarters of the
plot. False attribution regions correspond to the top-left (prolate-platy) and
bottom-right (oblate-acicular) regions. In other words, the curves with
positive slope refer to the true attribution, and the curves with negative
slopes refer to the false attribution.

\subsubsection{Oxygen-saturated phase LnBaMn$_{1.96}$Fe$_{0.04}$O$_{6}$}

In contrast to LnBaMn$_{1.96}$Fe$_{0.04}$O$_{5}$ the left-hand line of the
doublet turn to be more intense in the LnBaMn$_{1.96}$Fe$_{0.04}$O$_{6}$
series. The swap between the $\pm1/2\rightarrow\pm1/2$ and $\pm1/2\rightarrow
\pm3/2$ lines originates from the reversal of the sign of $V_{zz}.$Here again
the negative sign of $V_{zz}$ is consistent with the ionic point charge model,
in which only the charges in first coordination sphere are taken into
account\cite{HI}. The case $R_{Q}>1$ remains unchanged. The origin of
asymmetry is therefore attributable to either platy habits of the
crystallites, or to oblate vibrational ellipsoids. Both of them are expected
\textit{a priory} and their combined effect must be additive, but not partly
extinguishing each other as in above case of LnBaMn$_{1.96}$Fe$_{0.04}$O$_{5}%
$. The anisotropies $\langle x^{2}\rangle-\langle z^{2}\rangle$ indicated on
Fig.8 are the "effective" values given by the sum of true and equivalent to
the pair ($G_{1},G_{2}$) values.

The negative quadrupole splitting $\Delta E_{Q}$ decreases with decreasing the
size of Ln. This is in agreement with the ionic point charge model\cite{HI}.
The absolute value of EFG correlates with distortion $D_{6}$. In agreement
with abrupt change of distortion between Sm and Nd as shown in Fig.3 the
absolute value of $\Delta E_{Q}$ abruptly drops from 0.7 mm/s to 0.2 mm/s as
the Ln size increases from Sm to Nd. The oxygen-saturated "O$_{6}$"-family can
be thus divided into two subseries, according to their values of $\Delta
E_{Q}$. The spectra of small-size and large size Ln-subfamilies are shown in
Figs. 8 and 9, respectively. This observation of two subfamilies agrees well
with two types of behavior in the phase diagram reported
previously\cite{NKIOYU, Aka}. In the subfamily with small size Ln (Y, Gd, Sm)
the COO is formed by the orbital stacked in sequence aabb along c-axis, and in
the the large-size Ln-subfamily (Nd, Pr), the COO is not observed, while at
lowering temperature the ferromagnetism first sets in at $T_{\text{C}}$,
followed by the onset at $T_{\text{N}}$%
$<$%
$T_{\text{C}}$ of antiferromagnetic order of so-called A-type\cite{NKIOYU,
Aka}. 

\subsubsection{Cation-disordered phase Ln$_{0.5}$Ba$_{0.5}$Mn$_{0.98}%
$Fe$_{0.02}$O$_{3}$}

M\"{o}ssbauer spectra for the cation-disordered phase Ln$_{0.5}$Ba$_{0.5}%
$Mn$_{0.98}$Fe$_{0.02}$O$_{3}$ are shown in Fig.10. Iron is in octahedral
coordination wherein the small quadrupole splitting can be associated either
with the random short-range strain in oxygen sublattice or with the randomness
in the (Ba,La)-coordination sphere of the $^{57}$Fe nuclei.%
\begin{figure}
[ptb]
\begin{center}
\includegraphics[
height=4.2367in,
width=2.7994in
]%
{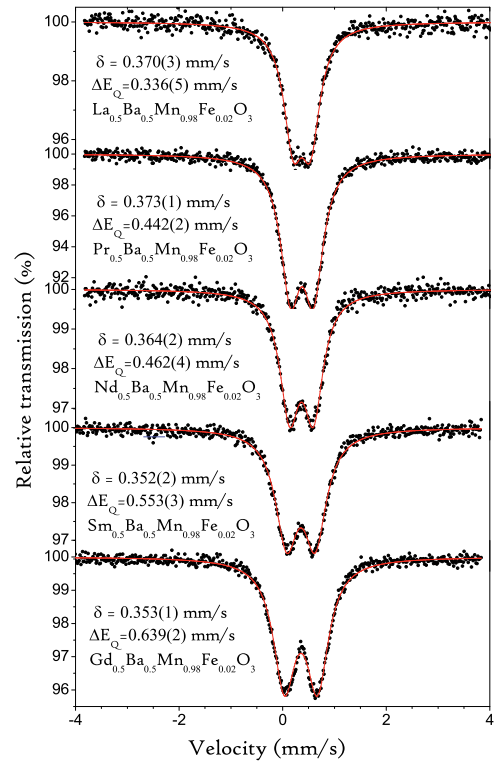}%
\caption{{}M\"{o}ssbauer spectra in disordered phase Ln$_{0.5}$Ba$_{0.5}%
$Mn$_{0.98}$Fe$_{0.02}$O$_{3}.$Fitted values of the quadrupole splittings and
chemical shifts are indicated.}%
\label{f10}%
\end{center}
\end{figure}

\subsubsection{ M\"{o}ssbauer line intensities versus March-Dollase
parameters}

The obtained formulas (\ref{06}) and (\ref{07}) allow to interpret the
M\"{o}ssbauer line intensities in terms of the Rietveld profile parameters
$G_{1}$ and $G_{2}$. In order to compare the Rietveld and M\"{o}ssbauer
results we present first in Fig.11 the contour plot of the doublet asymmetries
$R$ in function of $G_{1}$ and $G_{2}$. Different values of $R$ are shown by
different colors. Total ranges of variation for the alignment degree is
$0<G_{1}<\infty$ and for the fraction of unoriented phase $0<G_{2}<1$. In
Fig.11 the ranges $0<G_{1}<2$ and $0<G_{2}<1$ are presented.%

\begin{figure}
[ptb]
\begin{center}
\includegraphics[
height=4.337in,
width=2.9542in
]%
{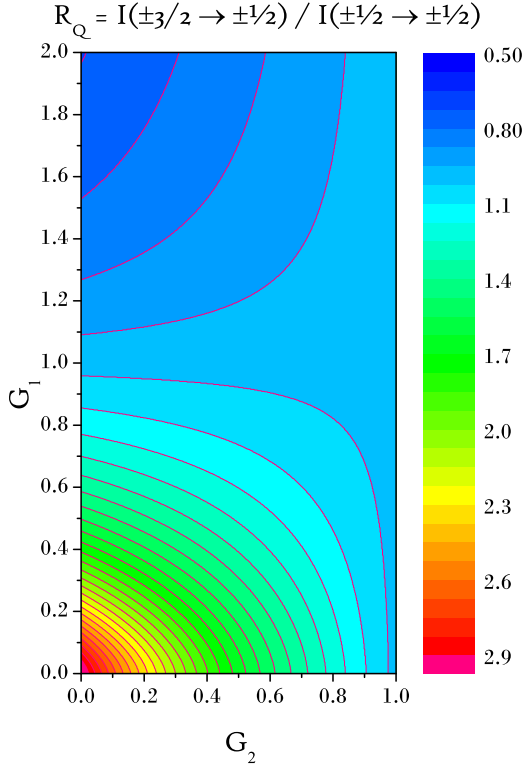}%
\caption{Contour plot of the M\"{o}ssbauer spectra asymmetry $R$ versus
$G_{1}$ and $G_{2}$ (Eqs. \ref{06} and \ref{07}). The upper half of the plot
corresponds to acicular habitus and the lower part of the plot correspond to
the platy habitus of the crystals. Note that only the region of relatively
weak degrees of alignment for acicular habitus is shown, while full range
$0<G_{1}<1$ is shown for platy habitus crystallites. The region of relatively
weak alignments exhibits symmetric dependence of $R$ on $G_{1}$ and $G_{2}$.}%
\label{f11}%
\end{center}
\end{figure}
%

\begin{figure}
[ptb]
\begin{center}
\includegraphics[
height=4.4477in,
width=2.7639in
]%
{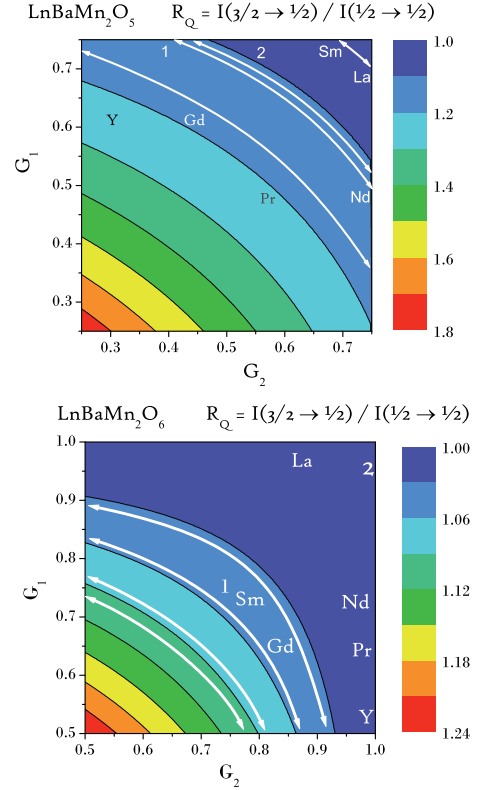}%
\caption{Regions of $R$ vs. $G_{1}$,$G_{2}$ plot corresponding to our samples
of GdBaMn$_{1.96}$Fe$_{0.04}$O$_{5}$ (top panel) and GdBaMn$_{1.96}$%
Fe$_{0.04}$O$_{6}$(bottom panel). The arcs show the spectral asymmetries from
Figs. 5 and 8 and the element symbols indicate the texturing coordinates
($G_{1}$,$G_{2}$) of the of the samples from the Table1.}%
\label{f12}%
\end{center}
\end{figure}

Since all the data in Table 1 indicate the platy habits with $0.25<G_{1}<0.75
$ in all our LnBaMn$_{1.96}$Fe$_{0.04}$O$_{5}$ samples, and with $0.5<G_{1}<1
$ we select these regions in Fig.12, (a) and (b), respectively. On the other
hand, both these regions correspond to the small enough experimental
asymmetries for the "O$_{5}$" and "O$_{6}$", respectively. The experimental
M\"{o}ssbauer doublet asymmetries from Figs. 5 and 8 are marked with the arc
segments, and the data from Table 1 are indicated by chemical symbol of Ln at
points with coordinates (G$_{1}$,G$_{2} $). Interestingly, it turns out that
the effects of $G_{1}$ and $G_{2}$ on the asymmetry turns out to be
approximately equal.

The detailed analysis of Fig.12 confirms the suggested above picture. In the
"O$_{5}$" series (Fig.12, top), the asymmetry indicated by symbols is stronger
than the asymmetry indicated by arcs. It means that GKE reduces the total
asymmetry indicated by arcs. In the "O$_{6}$" series (Fig.12, bottom), the
asymmetry indicated by arcs is stronger than the asymmetry indicated by
symbols. It means that GKE enhances the total asymmetry indicated by arcs.

\subsubsection{Theory of GKE in oriented powders}

The effects of texture and GKE on spectral asymmetry are only qualitatively
considered above to be additive or extinguishing. To express it more exactly,
the expression for intensity ratio was formulated\cite{Pfan1},%

\begin{equation}
R=\frac{\int_{0}^{\pi/2}M(\theta)(1+\cos^{2}\theta)e^{-\alpha\cos^{2}\theta
}d\theta}{\int_{0}^{\pi/2}M(\theta)(2/3+\sin^{2}\theta)e^{-\alpha\cos
^{2}\theta}d\theta} \label{10}%
\end{equation}
however, the solutions were yet found either for texture effects, or for GKE,
separately only. Substituting the MD function (Eq.1) for $M(\theta)$ we
propose a general solution for the combined effect of texture and vibrational
anisotropy. A replacement of $M(\theta,U,r)$ with the minimum texture function
\cite{Pfan2,Gren} makes the Eq. (10) integrable:%

\begin{equation}
R(\alpha)=\frac{(15V-6){}f(\alpha)+(12-15V)g(\alpha)+h(\alpha)}%
{(25V-10)f(\alpha)+(28-45V)g(\alpha)-h(\alpha)} \label{11}%
\end{equation}
Here $V=V(G_{1},G_{2})$, $f(\alpha)={}_{1}F_{1}(\frac{1}{2},\frac{3}%
{2},-\alpha)$ is the Kummer confluent hypergeometric function, $g(\alpha
)={}(f(\alpha)-e^{-\alpha})/\alpha$ and $h(\alpha)=(30-45V)(3g(\alpha
)-2e^{-\alpha})/4\alpha.$ The minimum texture function (MTF) compatible with
the MD function was taken as follows:%
\begin{equation}
\text{MTF}(x)=\left(  \frac{15}{4}-\frac{45}{4}x^{2}\right)  V(G_{1}%
,G_{2})+\frac{15}{2}x^{2}-\frac{3}{2} \label{09}%
\end{equation}
with $x=\cos\theta.$ The solution (\ref{11}) gives the exact solution only for
small asymmetries, because MTF is only a good approximation for the MD
function for $R$ close to 1. A perfect approximation for the broad range of
asymmetries is given elsewhere\cite{RSUN}.

\subsection{Implications for the nuclear inelastic scattering (NIS)
spectroscopy\bigskip}

\subsubsection{Putting in use the Rietveld analysis for the NIS spectra on
synchrotron radiation}

One of the problems, in which the well-oriented powders of platy or acicular
crystallites can be useful is the problem of determination of the partial
phonon density of states of $^{57}$Fe in anisotropic materials using the NIS
spectroscopy\cite{Seto}. The vibrational density of states is conventionally
derived from measuring the NIS spectra in single crystals along different
crystal axis. This Section presents a proposal for the novel NIS experiments
using the oriented powder samples.

In the NIS spectrum of an anisotropic single crystal the phonon DOS is
weighted by the squared projection of the phonon polarization vectors to the
wave vector of the x-ray quantum\cite{KCR}. Three projected densities of
states $g_{\zeta}(E)$ ($\zeta=x,y,z$) of an anisotropic layered material can
be determined my measuring the single crystal spectra at three different
orientations. In the uniaxially anisotropic material, the powder-averaged NIS
spectrum is isotropic, however, measuring two spectra on a sample with
preferred orientation of crystallites, $W(\omega_{1})$ and $W(\omega_{2})$ at
different angles(Fig.13) provides the full basis for determination of both DOS
functions $g_{x}(E)$ and $g_{z}(E)$. For this purpose the texture of a sample
should be well enough characterized, using Rietveld analysis and M\"{o}ssbauer spectroscopy.

The simplest experiment that we propose for measuring two components of the
uniaxially anisotropic DOS on a oriented sample involves the sample stage
rotated around the horizontal axis perpendicular to the incident beam
(Fig.13). Therefore, it is worth to describe the DOS in terms of the angles
$\theta$, $\omega$ and the Rietveld MD function preferred orientation
parameters $G_{1}$ and $G_{2}$.%
\begin{figure}
[ptb]
\begin{center}
\includegraphics[
height=1.7115in,
width=1.5065in
]%
{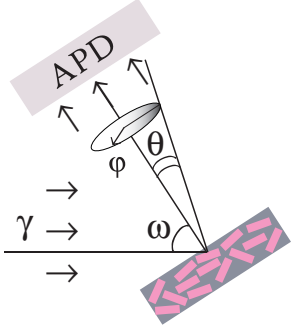}%
\caption{Geometry of nuclear inelastic scattering experiment. }%
\end{center}
\end{figure}

\subsubsection{Deriving the components of anisotropic vibrational DOS}

Let us launch the incident beam under the angle $\vartheta=\theta+\omega$ with
respect to the preferred axis (z-axis) of a platelike or a needle-like
crystallite. Each DOS component is weighted by the projection of the phonon
polarization vectors to the wave vector of the x-ray quantum, therefore, the
phonon DOS projected on the direction defined by the angle $\vartheta$ can be written:%

\begin{equation}
g_{E}(\vartheta)=g_{z}(E)\cos^{2}\vartheta+g_{x}(E)\sin^{2}\vartheta
\end{equation}
Powder averaging of the DOS consists in integrating these two terms\ with the
volume of crystallites $D(\vartheta,\phi)d\Omega$ whose z-axis lies within the
cone shell element $d\Omega$ and result in%

\begin{equation}
\left\langle g(E)\right\rangle =g_{x}(E)+\Delta g_{zx}(E)\int D(\vartheta
,\phi)\cos^{2}\vartheta d\Omega\label{03}%
\end{equation}
Here the normalization of the texture function to unity and the notation
$\Delta g_{zx}(E)=g_{z}(E)-g_{x}(E)$ are employed. The polar function
$M(\theta,U,r)$, independent on the azimuth angle, is to replace
$D(\vartheta,\phi)$ via the coordinate transform from the frame of the beam to
the frame of the rotation stage. The ratio of angular elements $d\Omega
_{\text{beam}}/d\Omega_{\text{stage}}$ is $\sin$ $\vartheta d\vartheta
d\phi/\sin\theta d\theta d\varphi$ and the Jacobian of this transform is
$\sin$ $\vartheta/\sin\theta$. Using%

\begin{equation}
\cos\vartheta=\cos\theta\cos\omega-\sin\theta\sin\omega\cos\varphi\label{55}%
\end{equation}
we obtain for the uniaxial symmetry%

\begin{equation}
\left\langle \cos^{2}\vartheta\right\rangle =\left\langle \cos^{2}%
\theta\right\rangle \cos^{2}\omega+%
\frac12
\left\langle \sin^{2}\theta\right\rangle \sin^{2}\omega\label{05}%
\end{equation}
Using the inegrated MD texture function (Eqs.\ref{06} and \ref{07}), we
substitute $\left\langle \sin^{2}\theta\right\rangle $ and $\left\langle
\cos^{2}\theta\right\rangle $ with $V(G_{1},G_{2})$ and $1-V(G_{1},G_{2})$ and
obtain
\begin{equation}
\left\langle \cos^{2}\vartheta\right\rangle =\left[  1-V(G_{1},G_{2})\right]
\cos^{2}\omega+\frac{V(G_{1},G_{2})}{2}\sin^{2}\omega\label{04}%
\end{equation}

From the Eq.(\ref{03}) a couple of measurements of DOS $g_{1}(E)$ and
$g_{2}(E)$ at the angles $\omega_{1}$ and $\omega_{2}$ leads immediately to
determination of both DOS components:%

\begin{equation}
\Delta g_{zx}(E)=\frac{\Delta g_{12}(E)}{\left[  1-\frac{3}{2}V(G_{1}%
,G_{2})\right]  (\cos^{2}\omega_{1}-\cos^{2}\omega_{2})}\text{ \ \ \ }%
\end{equation}

\begin{equation}
g_{x}(E)=g_{12}(E)-\frac{\Delta g_{12}(E)}{2}\frac{(\cos^{2}\omega_{1}%
+\cos^{2}\omega_{2})}{(\cos^{2}\omega_{1}-\cos^{2}\omega_{2})}\text{ \ \ \ }%
\end{equation}
where $g_{12}(E)=g_{1}(E)/2+g_{2}(E)/2$ and $\Delta g_{12}(E)=g_{1}%
(E)-g_{2}(E)$.

\section{Concluding Remarks}

Two aspects of this study were dealt with the methodological and material
issues. First, we proposed putting the Rietveld analysis in use for nuclear
inelastic scattering spectroscopy and developed the method of derivation of
the anisotropic phonon DOS from the experiments on oriented powder samples.

Second, in the material aspect of investigating the structure of doped
manganites we found a seemingly contradictory events of persistence of the
charge-orbital order observed in magnetization and in lattice cell dimension,
coexisting with the disordered-state single-site M\"{o}ssbauer spectra in most
of the samples, excluding LaBaMn$_{1.96}$Fe$_{0.04}$O$_{5}$, and, possibly,
SmBaMn$_{1.96}$Fe$_{0.04}$O$_{5}$. However, a plausible interpretation of this
combination of results is given suggesting the adaptiveness of the residual
short-range COO towards the quenched disorder related to the random
distribution of impurities.

In oxygen-saturated state ($y=6$) the manganites exhibit the charge and
orbital order at ambient temperature for Ln=Y, Gd, Sm, but unordered e$_{g}%
$-electronic system for Ln=La,Pr,Nd. Fourfold increase of quadrupole splitting
was observed in charge and orbitally ordered manganites compared to unordered
ones. This is in agreement with the jumplike increase of distortion of reduced
perovskite-like cell in the charge and orbitally ordered structures. The light
(i.e. with 2\%) substitution of Mn by Fe suppresses the temperatures of
structural and magnetic transitions by 20 to 50 K.

\section{Acknowledgements}

The authors acknowledge the financial support provided through joint JSPS-RFBR
Grant 07-02-91201. The work at School of Engineering was additionally
supported by Asahi Glass Foundation.

\section{Figure Captions}

Fig.1. The crystal structures and symmetry groups of the layer-ordered
LnBaMn$_{2}$O$_{5}$(a), LnBaMn$_{2}$O$_{6}$(b) and disordered Ln$_{0.5}%
$Ba$_{0.5}$MnO$_{3}$ (c).

Fig.2. Powder x-ray diffraction patterns of the tetragonal phases of
GdBaMn$_{1.96}$Fe$_{0.04}$O$_{5}$ and GdBaMn$_{1.96}$Fe$_{0.04}$O$_{6}$
refined using space groups $P4/nmm$ (No. 129) and $P4/mmm$ (No.123), respectively.

Fig.3. Lattice parameters of the reduced perovskite-like cell vs. volume of
this cell. Sm$_{0.9}$Nd$_{0.1}$BaMn$_{1.96}$Fe$_{0.04}$O$_{x}$ and Sm$_{0.1}%
$Nd$_{0.9}$BaMn$_{1.96}$Fe$_{0.04}$O$_{x}$ are denoted by "1" and "2", respectively.

Fig.4.Magnetic susceptibility $M/H$ measured in the external field $H$ of 1
kOe per mole of formula units in LnBaMn$_{1.96}$Fe$_{0.04}$O$_{6}$ for Ln=Y,
Sm and (Nd$_{0.9}$Sm$_{0.1}$). The zero-fied-cooled magnetization was measured
at heating the samples up to $T_{\text{max}}$ of 600 K (Ln=Y), 400 K (Ln=Sm)
and 370 K (Ln=Nd$_{0.9}$Sm$_{0.1}$) and then at cooling from $T=T_{\text{max}%
}$. The arrows indicate the temperatures of phase transitions in corresponding
pure LnBaMn$_{2}$O$_{6}$ without Fe substitution as reported
previously\cite{NKIOYU,NYU}.

Fig.5. Mossbauer spectra in oriented samples of $^{57}$Fe-doped oxygen-poor
series of the A-site ordered layered manganites LnBaMn$_{1.96}$Fe$_{0.04}%
$O$_{5}$. Fitted values of isomer shift, quadrupole splitting, and area
asymmetry are indicated on the spectra. The spectra are fitted with one or two
asymmetric doublets. Meltdown of Mn$^{2+}$/Mn$^{3+}$charge order results in
single-site spectra for $^{57}$Fe, while residue of the second doublet is
attributed to the remainder of the charge/orbital order.

Fig.6. Determination of the parameters of texture $G_{1}$ and vibrational
anisotropy $\alpha$ starting from the ratio of line intensities of
M\"{o}ssbauer doublet. Either texture or GKE is supposed to be a single source
of spectral asymmetry. Strength of preferred orientation $G_{1}$ is plotted
vs. spectral asymmetry $R_{Q}=I_{\pm1/2\rightarrow\pm3/2}/I_{\pm
1/2\rightarrow\pm1/2}$ for four values of the fraction of unoriented phase
$G_{2}$.

Fig.7. The equivalent vibrational anisotropy parameter $\alpha$ versus degree
of alignment $G_{1}\ $at four values of the unoriented phase $G_{2}$. The
curves with positive slope refer to the true assignment, and the curves with
negative slopes refer to the false assignment of the doublet lines to the
$\pm1/2\rightarrow\pm3/2$ and $\pm1/2\rightarrow\pm1/2$ transitions.

Fig.8. Mossbauer spectra in oriented samples of $^{57}$Fe-doped oxygen-rich
series of the charge and orbitally ordered layered manganites LnBaMn$_{1.96}%
$Fe$_{0.04}$O$_{6}$. The spectra are fitted with one asymmetric doublet.
Meltdown of Mn$^{3+}$/Mn$^{4+}$charge order results in single-site spectra for
$^{57}$Fe. Fitted values of isomer shift, quadrupole splitting, and area
asymmetry are indicated on the spectra. The values of $\langle x^{2}%
\rangle-\langle z^{2}\rangle$ are the "effective" anisotropies given by the
sum of true and equivalent to the pair ($G_{1},G_{2}$) values.

Fig.9. Mossbauer spectra in the large-size-Ln subseries of the oxygen-rich
series of the A-site ordered layered manganites LnBaMn$_{1.96}$Fe$_{0.04}%
$O$_{6}$ showing no COO transition. The spectra are fitted with one symmetric doublet.

Fig.10. M\"{o}ssbauer spectra in disordered phase Ln$_{0.5}$Ba$_{0.5}%
$Mn$_{0.98}$Fe$_{0.02}$O$_{3}.$Fitted values of the quadrupole splittings and
chemical shifts are indicated.

Fig.11. Contour plot of the M\"{o}ssbauer spectra asymmetry $R$ versus $G_{1}
$ and $G_{2}$ (Eqs. \ref{06} and \ref{07}). The upper half of the plot
corresponds to acicular habitus and the lower part of the plot correspond to
the platy habitus of the crystals. Note that only the region of relatively
weak degrees of alignment for acicular habitus is shown, while full range
$0<G_{1}<1$ is shown for platy habitus crystallites. The region of relatively
weak alignments exhibits symmetric dependence of $R$ on $G_{1}$ and $G_{2}$.

Fig.12. Regions of $R$ vs. $G_{1}$,$G_{2}$ plot corresponding to our samples
of GdBaMn$_{1.96}$Fe$_{0.04}$O$_{5}$ (top panel) and GdBaMn$_{1.96}$%
Fe$_{0.04}$O$_{6}$(bottom panel). The arcs show the spectral asymmetries from
Figs. 5 and 8 and the element symbols indicate the texturing coordinates
($G_{1}$,$G_{2}$) of the samples from the Table1.

Fig. 13. Geometry of nuclear inelastic scattering experiment.


\section{References}

\begin{thebibliography}{99}                                                                                               %


\bibitem {NKIOYU}T. Nakajima, H. Kageyama, M. Ichihara, K. Ohoyama, H.
Yoshizawa, and Y. Ueda, J. Solid St. Chem. \textbf{177}, 987 (2004).

\bibitem {WAR}A.J. Williams, J.P. Attfield, and S.A.T. Redfern, Phys. Rev.
\textbf{B} \textbf{72}, 184426 (2005).

\bibitem {EPL}A. I. Rykov, Europhys. Lett. \textbf{85} (2009) 16003.

\bibitem {MCDRS}J.P. Chapman, J.P. Attfield, M. Molgg, C.M. Friend, and T.P.
Beales, Angew. Chem. Int. Ed. Engl. \textbf{35}, 2482 (1996).

\bibitem {MA}J.A. McAlister, J.P. Attfield, J. mater. Chem. \textbf{8}, 1291 (1998).

\bibitem {Millange}F. Millange, V. Caignaert, B. Domeng\`{e}s, and B. Raveau,
Chem. Mater. \textbf{10}, 1974 (1998).

\bibitem {NKYOU}T. Nakajima, H. Kageyama, H. Yoshizawa, K. Ohoyama, and Y.
Ueda, J. Phys. Soc. Jpn. \textbf{72}, 3237 (2003).

\bibitem {NYU}T. Nakajima, H. Yoshizawa and Y. Ueda, J. Phys. Soc. Jpn.,
\textbf{73}, 2283 (2004).

\bibitem {Aka}D. Akahoshi, Y. Okimoto, M. Kubota, R. Kumai, T.Arima, Y.
Tomioka, and Y. Tokura, Phys. Rev. B 70, 064418 (2004).

\bibitem {NKU}T. Nakajima, H. Kageyama, Y. Ueda, J. Phys. Chem. Solids,
\textbf{63}, 913 (2002).

\bibitem {RCR}A. Rykov, V. Caignaert, and B. Raveau, J. Solid. St. Chem.
\textbf{109}, 295 (1994).

\bibitem {KCR}V.G. Kohn, A.I.\ Chumakov, R. R\"{u}ffer, Phys. Rev. B.
\textbf{58}, 8437 (1998).

\bibitem {DBW}R.A. Young and D.B. Wiles, Adv. X-ray. Anal. \textbf{24}, 1, (1981).

\bibitem {Full}J. Rodriguez-Carvajal. Physica \textbf{B 192}, 55 (1993).

\bibitem {MSCB}F. Millange, E. Suard, V. Caignaert, and B. Raveau, Mat. Res.
Bull. \textbf{34}, 1 (1999).

\bibitem {Caignaert}V. Caignaert, F. Millange, B. Domeng\`{e}s, and B. Raveau,
Chem. Mater. \textbf{11}, 930 (1999).

\bibitem {Er}L. Er-Rakho, C. Michel, P. Lacorre and B. Raveau, J. Solid State
Chem. \textbf{73} (1988) 531.

\bibitem {JPSJ2002}T. Nakajima, H. Kageyama, H. Yoshizawa, and Y. Ueda, J.
Phys. Soc. Jpn \textbf{71}, 2843 (2002)

\bibitem {Akah}D. Akahoshi, M. Uchida, Y. Tomioka, T. Arima, Y. Matsui, and Y.
Tokura, Phys. Rev. Lett. \textbf{90}, 177203 (2003).

\bibitem {UN}Y. Ueda and T. Nakajima, Progr. Solid St. Chem. \textbf{35}, 397 (2007).

\bibitem {Belik}A.A. Belik, N. Hayashi, M. Azuma, S. Muranaka., M. Takano, and
E. Takayama-Muromachi, J. Solid St. Chem. \textbf{180}, 3401 (2007).

\bibitem {inc}Incorporated into FULLPROF program, http://www-llb.cea.fr/fullweb/powder.htm.

\bibitem {Dollase}W.A. Dollase, J. Appl. Cryst. \textbf{19} (1986) 267-272.

\bibitem {Riet}H.M. Rietveld, J. Appl. Cryst. \textbf{2,} 65 (1969).

\bibitem {Taskin}A.A. Taskin,A.N. Lavrov, Yoichi Ando, Progr. Solid St. Chem.
\textbf{35} 481 (2007).

\bibitem {parallel}A.I. Rykov, Y. Ueda and K. Nomura, arXiv:0902.2027v1 [cond-mat.str-el].

\bibitem {Sh}G.K. Shenoy, G.M. Friedt, Nucl. Inst. Meth. \textbf{136}, 569 (1976).

\bibitem {HI}A.I. Rykov, A. Ducouret, N. Nguyen, V. Caignaert, F. Studer and
B. Raveau, Hyperfine Interact. \textbf{77} (1993) 277.

\bibitem {Mom}A.W. Mombr\'{u}, K. Prassides, C. Christides, R.Erwin, M.
Pissas, C. Mitros, and D. Niarchos, J. Phys.: Cond. Mat. \textbf{10}, 1247 (1998).

\bibitem {Pfan1}H.-D. Pfannes and U. Gonser, Appl. Phys. \textbf{1}, 93 (1973).

\bibitem {Pfan2}H.-D. Pfannes and H. Fisher, Appl. Phys. \textbf{13}, 317 (1977).

\bibitem {Gren}J.-M. Greneche and F. Varret, J. Phys. \textbf{C15}, 5333 (1982).

\bibitem {RSUN}A.I. Rykov, M. Seto, Y. Ueda and K. Nomura, arXiv:0902.1801v1 [cond-mat.mtrl-sci]

\bibitem {Seto}M. Seto, Y. Yoda , S. Kikuta , X.W. Zhang and M. Ando, Phys.
Rev. Lett.\textbf{74}, 3828 (1995).
\end{thebibliography}

\begin{thebibliography}{9999}                                                                                             %

\end{thebibliography}
\end{document}